\documentclass[11pt,a4paper]{article}

\usepackage{latexsym,bezier,enumerate,dcolumn,color}
\usepackage{amsmath,amsthm,amsopn,amstext,amscd,amsfonts,amssymb}

\usepackage[round]{natbib}
\usepackage[]{graphicx}
\usepackage[affil-it]{authblk}
\usepackage{geometry}
\usepackage[T1]{fontenc}
\usepackage[latin9]{inputenc}
\usepackage{multicol,multirow,setspace}
\usepackage{caption}
\usepackage{booktabs}

\usepackage{tikz, pgfplots}
\pgfplotsset{compat=1.8}
\usetikzlibrary{patterns}

\usepackage{setspace}
\usepackage{hyperref}

\newif\ifblinded
\blindedtrue

\begin{document}

\title{The Probabilistic Final Standing Calculator: a fair stochastic tool to handle abruptly stopped football seasons}

\date{}


\author{Hans Van Eetvelde}
\affil{Department of Applied Mathematics, Computer Science and Statistics, Ghent University, Ghent, Belgium}

\author{Lars Magnus Hvattum}
\affil{Faculty of Logistics, Molde University College, Norway}

\author{Christophe Ley}
\affil{Department of Applied Mathematics, Computer Science and Statistics, Ghent University, Ghent, Belgium}



\maketitle


%

%
%
%
%
%
%


\begin{abstract}

The COVID-19 pandemic has left its marks in the sports world, forcing the full-stop of all sports-related activities in the first half of 2020. Football leagues were suddenly stopped and each country was hesitating between a relaunch of the competition and a premature ending. Some opted for the latter option, and took as the final standing of the season the ranking from the moment the competition got interrupted. This decision has been perceived as unfair, especially by those teams who had remaining matches against easier opponents. In this paper, we introduce a tool to calculate in a fairer way the final standings of domestic leagues that have to stop prematurely: our Probabilistic Final Standing Calculator (PFSC). It is based on a stochastic model taking into account the results of the matches played and simulating the remaining matches, yielding the probabilities for the various possible final rankings. We have compared our PFSC with state-of-the-art prediction models, using previous seasons which we pretend to stop at different points in time. We illustrate our PFSC by showing how a  probabilistic ranking of the French Ligue 1 in the stopped 2019-2020 season could have led to alternative, potentially fairer, decisions on the final standing.

\end{abstract}



\newpage
\section{Introduction}

The COVID-19 pandemic has left its marks in the world of sports, forcing in February-March 2020 the full-stop of nearly  all sports-related activities around the globe. This implied that the national and international sports leagues were abruptly stopped, which had dramatic impacts especially in the  most popular sport, football (or soccer). Indeed, the absence of regular income through ticket selling, television money and merchandising around live matches entailed that a large majority of professional clubs no longer were able to pay their players and other employees \citep{sky2020, kicker2020}. Professional football having become such a big business, this also had an impact on other people whose main income was related to football matches \citep{Arsenal}. And, last but not least, the fans were starving to have their favourite sport resume. However, the very intense and uncertain times made it a tough choice for decision-makers to take the risk of letting competitions go on, especially as it would be without stadium attendance for the first weeks or possibly months. Consequently, some leagues never resumed. The Dutch Eredivisie was the first to declare that it would not be continued, on April 25 2020, followed by the French Ligue 1 on April 28 and the Belgian Jupiler Pro League on May 15. The German Bundesliga was the first professional league to get restarted, followed by the English Premier League, Spanish Primera Division and Italian Serie A. 

The leagues that decided to not resume playing were facing a tough question: how to evaluate the current season? Should it simply be declared void, should the current ranking be taken as final ranking, or should only a part of the season be evaluated? Either decision would have an enormous financial impact given the huge amount of money at stake. Which team would be declared champion? Which teams would be allowed to represent their country at the European level and hence earn a lot of money through these international games, especially in the Champions League? Which teams would be relegated and, as a potential consequence, be forced to dismiss several employees? Moreover, the broadcasting revenue is  allocated on the basis of final standings. Different countries reacted differently: the Eredivisie had no champion nor relegated teams, but the teams qualified for the European competitions were decided based on the ranking of March 8. The Ligue 1 and the Jupiler Pro League, on the other hand, declared also a champion and relegated teams on the basis of the ranking from the moment their seasons got stopped. However, the Jupiler Pro League had to come back on this decision and the relegation was nullified. The Ligue 1 based their final standing on the ratio of points earned per game, since not all teams had played an equal amount of games when the season got stopped. Obviously, several teams were not pleased by such decisions, considering them to be unfair \citep{Holroyd2020} because, inevitably, this favoured those teams that had still to play the strongest opponents in the remaining matches over those that were looking forward to a rather light end-of-season. 

This naturally raises the question to find a more balanced, scientifically sound way to evaluate the final standing of abruptly ended seasons. It also represents a bigger challenge than evaluating in a fair way an abruptly stopped single game, where for instance cricket has adopted a mathematical formula known as the Duckworth-Lewis-Stern method \citep{DLSmethod}. 
The literature addressing the challenging question of how a stopped season should be evaluated in the most objective way is  fuelled by proposals from after the outbreak of COVID-19. The first and third authors of this paper used the current-strength based rankings by \citet{LeWiVa19}, a statistical model based on the bivariate Poisson distribution and weighted maximum likelihood, to provide all possible final standings of the Belgian Jupiler Pro League together with their associated probabilities. Their results were summarized in a mainstream journal \citep{coronaBelgie}. \citet{Guyon2020} proposed an Elo-based approach applied to the English Premier League. \citet{LaSp20} suggested an eigenvector-based approach, while \citet{csato2020coronavirus} discussed general criteria that a fair ranking should fulfil in such a situation and proposed the generalized row sum method. Recently, \cite{go20} used a statistical model to determine a ranking based on the expected total number of points per team. 

In this paper, we investigate the extent to which a relatively simple stochastic model can serve the purpose of producing fair final standings for prematurely stopped round-robin type football competitions. Our original approach   goes as follows. We construct a stochastic soccer model that is fitted on the played matches and then is used to simulate the remainder of the competition a large number of times, thus yielding for every team the probabilities to reach each possible final rank. This output is much richer in terms of information than giving only the most likely  or the expected ranking. This also explains the terminology for our model, namely Probabilistic Final Standing Calculator, which we abbreviate as PFSC. In order to assess its predictive strength, we compare our PFSC with two benchmark prediction models, namely the best performing model of \citet{LeWiVa19}, that uses a similar stochastic model to estimate the current strength of a team based on its matches played in the last two years, and the plus-minus-ratings approach of \citet{PaHv19} that is based on the strengths of the players making up the teams. For each model, the probabilistic final standing of a not yet ended season is obtained by simulating the remaining matches 100,000 times, which gives us for every team the probabilities of reaching each possible place in the final standing. It is not appropriate to compare the predictions of these models on the 2019-2020 competitions which were resumed after the break, since those matches were played under different circumstances, including the absence of fans. It has been shown \citep{Fischer2020} that these changed conditions could influence team performances, by lowering the effect of the home advantage. Therefore we rather compare the three models on the basis of the three preceding seasons of the five most important European football leagues (England, Spain, Germany, Italy and France), which we stopped artificially after every match day. Our evaluation of the models' performance is done in two ways: by means of the Rank Probability Score (RPS) \citep{epstein1969scoring} and the Tournament Rank Probability Score (TRPS) \citep{ekstrom2019evaluating}, see Section~\ref{sec:eval} for their definition. From this comparison, we can see at which point in time the PFSC is able to catch up with the two high-performing but more complicated prediction models. The reader may now wonder why we do not use any of these more elaborate models as PFSC; the reason is that we wish to propose a handy tool that sports professionals can indeed use without the need of too long computation time or large data collections. In the same vein, we will also make the PFSC  freely available on the software $\mathtt{R}$ \citep{R2020}.

The remainder of the paper is organized as follows. In Section~\ref{sec:methods} we describe our PFSC along with the two alternative models, as well as the two model performance evaluation metrics. Section~\ref{sec:results} then presents the results of this broad comparison. In Section~\ref{sec:France}, we illustrate the advantages of our PFSC by analyzing the French Ligue 1 season 2019-2020 and discussing how fairer decisions could be obtained on the basis of our PFSC. We conclude the paper with final comments in Section~\ref{sec:final}.

%
%
%

\section{Methods}
\label{sec:methods}

In this section, we start by explaining the PFSC (Section~\ref{sec:bPois}) and the two benchmark models (Sections~\ref{sec:ranks} and \ref{sec:pm_ratings}), followed by a description of the two evaluation measures for comparison (Section~\ref{sec:eval}). In what follows, we suppose to have a total of $n$ teams competing in a round-robin type tournament of $M$ matches.

\subsection{The PFSC: a bivariate Poisson-based model}
\label{sec:bPois}

For modelling football matches, the PFSC will make use of the bivariate Poisson distribution. Building on the original idea of \cite{Ma82} to model football match outcomes via Poisson distributions, the bivariate Poisson distribution has been popularized by \cite{KaNt03}.
Let $Y_{ijm}$ stand for the number of goals scored by team $i$ against team $j$ ($i,j\in \{1,...,n\}$) in match $m$ (where $m \in \{1,...,M\}$) and let $\lambda_{ijm}\geq0$ resp. $\lambda_{jim}\geq0$ be the expected number of goals for team $i$ resp. $j$ in this match. The joint probability function of the home and away score is then given by the bivariate Poisson probability mass function  
\begin{align*} 
&{\rm P}(Y_{ijm}=x, Y_{jim}=y) = \frac{\lambda_{ijm}^x \lambda_{jim}^y}{x!y!} \exp(-(\lambda_{ijm}+\lambda_{jim}+\lambda_{C})) \sum_{k=0}^{\min(x,y)} \binom{x}{k} \binom{y}{k}k!\left(\frac{\lambda_{C}}{\lambda_{ijm}\lambda_{jim}}\right)^k, \label{bivpoissonDens}
\end{align*}
where  $\lambda_{C}\geq0$ is a covariance parameter representing the interaction between both teams. This parameter is kept constant over all matches, as suggested in \cite{LeWiVa19}, who mentioned that models where this parameter depends on the teams at play perform worse. Note that $\lambda_{C}=0$ yields the Independent Poisson model. The expected goals $\lambda_{ijm}$ are expressed in terms of the strengths of team~$i$ and team~$j$, which we denote $r_i$ and $r_j$, respectively, in the following way: $\log(\lambda_{ijm})=\beta_0 + ({r}_{i}-{r}_{j})+h\cdot \rm{I}(\mbox{team $i$ playing at home})$, where $h$ is a real-valued parameter representing the home effect and is only added if team~$i$ plays at home, and $\beta_0$ is a real-valued intercept indicating the expected number of goals $e^{\beta_0}$ if both teams are equally strong and play on a neutral ground. The strengths $r_1,\ldots,r_n$ can take both positive and negative real values and are subject to the identification constraint $\sum_{i=1}^nr_i=0$. Over a period of $M$ matches (which are assumed to be independent), this leads to the likelihood function
\begin{equation}\label{lik}
L = \prod_{m=1}^{M}{\rm P}(Y_{ijm}=y_{ijm}, Y_{jim}=y_{jim}), 
\end{equation}
where $y_{ijm}$ and $y_{jim}$ stand for the actual number of goals scored by teams $i$ and $j$ in match $m$. The unknown values of the strength parameters $r_1,\ldots,r_n$ are then computed numerically as maximum likelihood estimates, that is, in such a way that they best fit a set of observed match results.

\cite{LeWiVa19} established that the bivariate Poisson model and its Independent counterpart are the best-performing maximum likelihood based models for predicting football matches. We refer the interested reader to Section 2 of \cite{LeWiVa19} for more variants of the bivariate Poisson model, as well as for Bradley-Terry type models where the outcome (win/draw/loss) is modelled directly instead of as a function of the goals scored. 

Using the bivariate Poisson model in the final standing prediction works as follows. The parameters $\lambda_C$, $\beta_0$, $h$ and the strength parameters $r_1,...,r_n$ are estimated using the matches played so far in the current season. Next, these parameters are used to simulate 100,000 times the remaining matches, by sampling the number of goals for each team in each match from the corresponding bivariate Poisson distribution. For each simulated end of season, a final standing is created based on the played and simulated matches, taking into account the specific rules of the leagues. The probabilistic final standing is then calculated by averaging the results over all 100,000 simulations, giving each team a probability to reach every possible rank. Note that \cite{go20} also used the bivariate Poisson distribution as their statistical model, but they only calculate expected ranks and not the complete probabilistic picture as we do.

This model is relatively simple compared to the benchmark models that we  describe below, but it has some nice properties that make it perfectly suited for determining the final standing of a prematurely stopped competition. First, the PFSC only takes into account match results, so data requirements are benign. Second, the PFSC only takes into account matches of the current season, so there is no bias to teams that performed well in the previous season(s). Third, each played game has the same weight in the estimation of the team strengths. These three properties make this method a fair way to evaluate an unfinished football season. On top of this, the code for the model can easily be executed in a short time.

\subsection{Current-strength based team ratings}\label{sec:ranks}
The first benchmark model is an extension of the previous model. The idea of \cite{LeWiVa19} was to use a weighted maximum likelihood, where the weight is a time depreciation factor $w_{time,m}>0$ for match $m$, resulting in
\begin{equation*}\label{lik2}
L = \prod_{m=1}^{M}\left({\rm P}(Y_{ijm}=y_{ijm}, Y_{jim}=y_{jim})\right)^{w_{time,m}}.
\end{equation*}
The exponentially decreasing time decay function is defined as follows: a match  played $x_m$ days back gets a weight of 
\begin{equation*}\label{smoother}
w_{time,m}(x_m) = \left(\frac{1}{2}\right)^{\frac{x_m}{\mbox{\small Half period}}}.
\end{equation*}
In other words, a match played \emph{Half period} days ago only contributes half as much as a match played today and a match played $3\times$\emph{Half period} days ago contributes 12.5\% of a match played today. This weighting scheme gives more importance to recent matches  and leads to a so-called current-strength ranking based on the estimated strength parameters of the teams. 

Another difference is that this model uses two years of past matches to estimate the team strengths. The half period is set to 390 days, as this was found to be the optimal half period by \cite{LeWiVa19} when evaluated on 10 seasons of the Premier League. The predicted probabilities for each rank in the final standing are obtained in the same way as in the PFSC.

\subsection{Plus-minus ratings}\label{sec:pm_ratings}

Plus-minus ratings, the second benchmark model, are based on the idea of distributing credit for the performance of a team onto the players of the team. We consider the variant of plus-minus proposed by \citet{PaHv19}. Each football match is partitioned into segments of time, with new segments starting whenever a player is sent off or a substitution is made. For each segment, the set of players appearing on the pitch does not change, and a goal difference is observed from the perspective of the home team, equal to the number of goals scored by the home team during the segment minus the number of goals scored by the away team. The main principle of the plus-minus ratings considered is to find ratings such that the sum of the player ratings of the home team minus the sum of the player ratings of the away team is as close as possible to the observed goal difference.

Let $S$ be the set of segments, $P_{h(s)}$ respectively $P_{a(s)}$  the set of players on the pitch for the home respectively away team during segment $s \in S$. Denote by $g(s)$ the goal difference in the segment as seen from the perspective of the home team. If a real-valued parameter $\beta_j$ is used to denote the rating of player $j$, the identification of ratings can be expressed as minimizing
\begin{align}
 \sum_{s \in S} \left( \sum_{j \in P_{h(s)}} \beta_j - \sum_{j \in P_{a(s)}} \beta_j - g(s) \right)^2, \nonumber
\end{align}
the squared difference between observed goal differences and goal differences implied by the ratings of players involved. To derive more reasonable player ratings, \citet{PaHv19} considered a range of additional factors, which we also consider here: 1) Segments have different durations, so the ratings in each segment are scaled to correspond with the length of the segment. 2) The home team has a natural advantage, which is added using a separate parameter. 
3) Some segments have missing players, either due to players being sent off by the referee or due to injuries happening after all allowed substitutions have been made. These situations are represented using additional variables corresponding to the missing players, while remaining player ratings are scaled so that their sum corresponds to an average rating for a full team. 4) The player ratings are not assumed to be constant over the whole data set, but rather to follow a curve that is a function of the age of players. This curve is modelled as a piece-wise linear function which is estimated together with the ratings by introducing corresponding age adjustment variables. 5) Each segment is further weighted by factors that depend on the recency of the segment and the game state. A complete mathematical formulation of the plus-minus rating system was provided by \citet{PaHv19}.

To move from plus-minus player ratings to match predictions, an ordered logit regression model is used. This model derives probabilities for a home win, a draw, and an away win based on a single value associated with each match: the sum of the ratings of the players in the starting line-up for the home team, minus the sum of the ratings of the players in the starting line-up for the away team, plus the home field advantage of the corresponding competition. 

As with the previous benchmark, the remaining matches of a league are simulated. However, some slight differences can be observed. Following \citet{SaHv19}, the starting line-ups of the teams are also simulated, based on the players available in the squads. Each player has a 10 \% chance of being injured or otherwise inadmissible for a given match. Subject to these random unavailable players, the best possible starting line-up is found, consisting of exactly one goalkeeper, and at least three defenders, three midfielders, and one forward. Based on this, probabilities of a home win, draw and away win are derived using the ordered logit regression model. Since this does not provide a goal difference, but just a result, the simulation further assumes that losing teams score 0 goals and drawing teams score one goal, whereas the number of goals for winning teams is selected at random from 1 to 3.

\subsection{Metrics to evaluate and compare the three models}
\label{sec:eval}

We have provided three proposals for predicting the final standings of abruptly stopped football seasons. These are evaluated by predicting, for several completed seasons from different top leagues, the remaining matches after artificially stopping each season after every match. The evaluation of their predictive abilities is done at two levels: single match outcomes and final season standings. For the former, we use the Rank Probability Score as metric, for the latter the Tournament Rank Probability Score.

The Rank Probability Score (RPS) is a proper scoring rule that preserves
the ordering of the ranks and places a smaller penalty on predictions that are closer
to the observed data than predictions that are further away from the observed data \citep{epstein1969scoring,gneiting2007strictly,CoFe12a}. The
RPS is defined for a single match as
$$
RPS =\frac{1}{R-1}\sum_{r=1}^R\left(\sum_{j=1}^r(o_j-x_j)\right)^2
$$
where $R$ is the number of possible outcomes, $o_j$  the empirical probability of outcome
$j$ (which is either 1 or 0), and $x_j$ the forecasted probability of outcome $j$. The smaller the RPS, the better the prediction. The RPS is similar to the Brier score, but measures the accuracy of a prediction differently when there are more than two ordered categories, by using the cumulative predictions in order to be sensitive to the distance. Let us give some further intuition about this metric. Let 1 stand for home win, 2 for draw and 3 for away win, so obviously $R=3$. The formula of the RPS can be simplified to 
$$
\frac{1}{2}\left((o_1-x_1)^2+(o_1+o_2-x_1-x_2)^2+(1-1)^2\right)=\frac{1}{2}\left((o_1-x_1)^2+(o_3-x_3)^2\right),
$$
which shows, for instance, that a home win predicted as draw is less severely penalized than would be a predicted away win in such case. 

\citet{ekstrom2019evaluating} extended the RPS to final tournament or league standings, and consequently termed it TRPS for Tournament RPS. The idea is very similar to the RPS, as the TRPS compares the  cumulative prediction $X_{rt}$ that team $t$ will reach at least rank $r$ (with lower values of $r$ signifying a better ranking) to  the corresponding empirical cumulative probability $O_{rt}$. The latter also only attains two different values: a column $t$ in $O_{rt}$ is 0 until the rank which team $t$ obtained in the tournament, after which it is 1. Consequently, the TRPS is defined as
$$
TRPS=\frac{1}{T}\sum_{t=1}^T\frac{1}{R-1}\sum_{r=1}^{R-1}(O_{rt}-X_{rt})^2,
$$
where $T$ is the number of teams and $R$ is the total number of possible ranks in a tournament or league. A perfect prediction will yield a TRPS of 0 while the TRPS increases when the prediction  worsens. The TRPS is a proper scoring rule, very flexible and handles partial rankings. It retains for league predictions the desirable properties of the RPS, and as such assigns lower values to predictions that are almost right than to predictions that are clearly wrong.



\section{Results}\label{sec:results}

The PFSC, the current-strength team ratings and the plus-minus player ratings are evaluated in terms of correctly predicting match outcomes and the final league table. The evaluation is conducted on the top leagues of England, France, Germany, Italy, and Spain, for the 2016-2017, 2017-2018 and 2018-2019 seasons. Each season and league is halted after each match day, and the outcomes of the remaining matches as well as the final league tables are predicted.

\begin{figure} 
\centering
\includegraphics[width=0.85\linewidth]{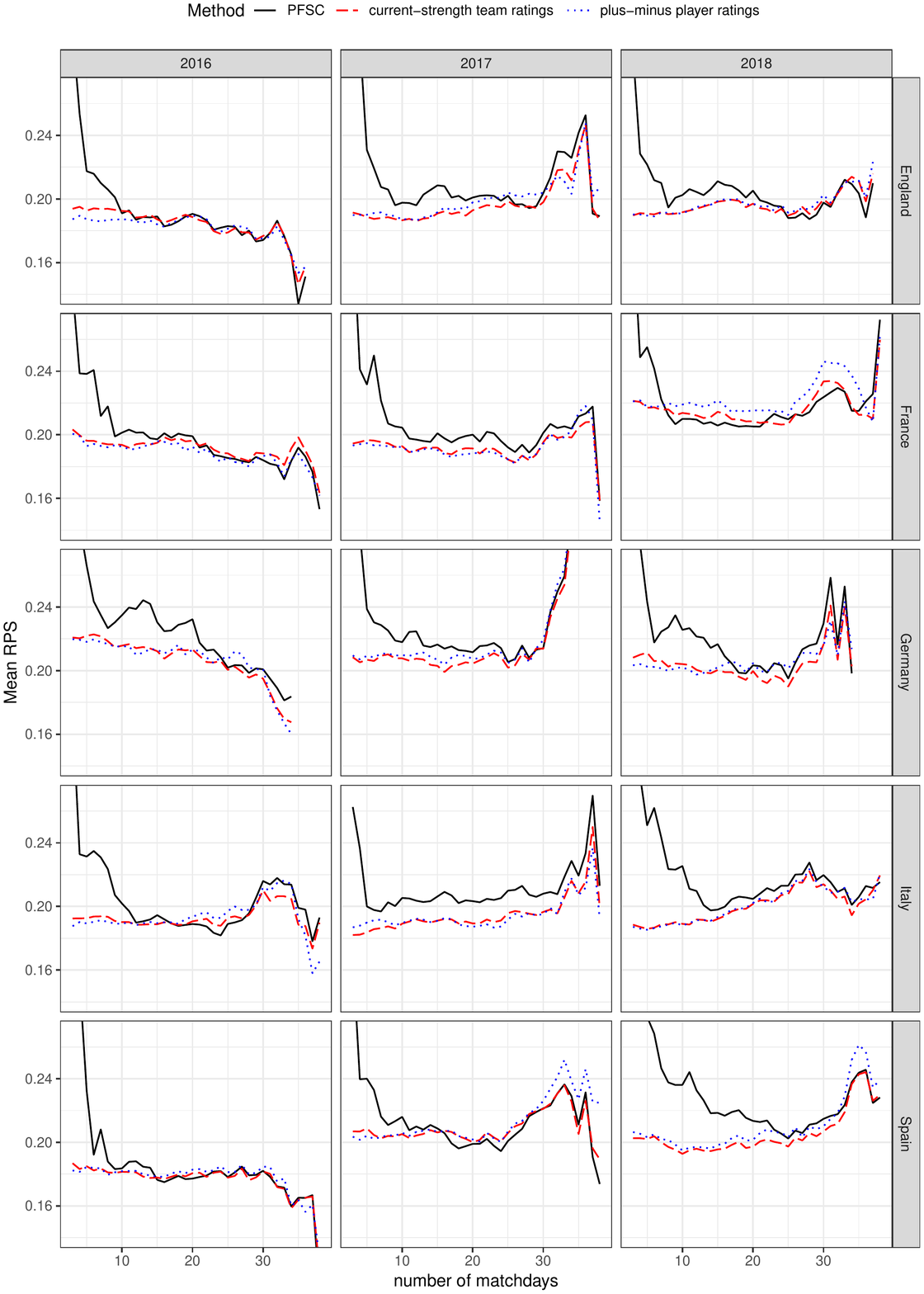}
  \caption{Mean RPS values calculated after each match day for five leagues and three seasons.}\label{fig:RPSValues}
\end{figure}

Figure~\ref{fig:RPSValues} shows the mean RPS of the remaining matches, given the current match day. The figure illustrates that the performances of the current-strength team ratings and the plus-minus player ratings are similar throughout. The PFSC is a much simpler model, and only uses data from the current season. Therefore, its performance is relatively bad in the beginning of the season. However, in most cases the PFSC converges towards the performance of the benchmark methods after around 10 match days, and in all but one of fifteen league-seasons it has caught up after 25 match days. The exception is the 2017-2018 season of the Italian Serie A, where the PFSC leads to worse predictions than the benchmarks throughout.

When nearing the end of the season, the RPS is calculated over few matches. Therefore, the mean RPS behaves erratically and sometimes increases sharply. This is because a single upset in one of the final rounds can have a large effect on the calculated RPS values. However, as the plots in Figure~\ref{fig:RPSValues} show, all three methods follow each other closely, indicating that the results that are difficult to predict are equally hard for all methods.

\begin{figure} 
\centering
\includegraphics[width=0.85\linewidth]{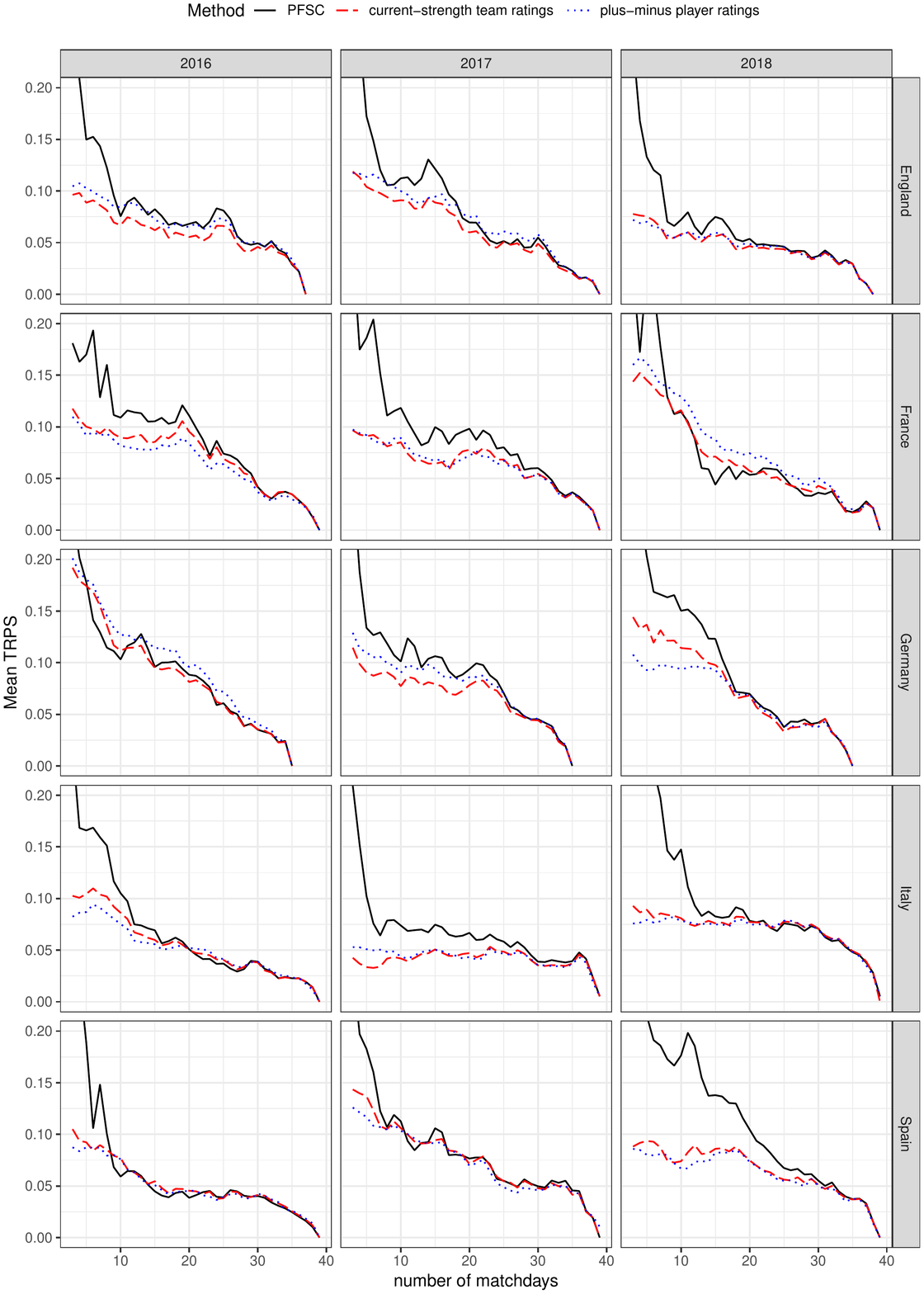}
  \caption{Mean TRPS values calculated after each match day for five leagues and three seasons.}\label{fig:TRSPValues}
\end{figure}

In Figure~\ref{fig:TRSPValues} the mean TRPS is shown for each league and season. As the final table becomes easier to predict the more matches have been played, the TRPS converges to zero. Therefore, the difference in performance among the three methods also converges to zero. However, the PFSC has a similar prediction quality as the benchmark methods already somewhere between 10 and 25 match days into the season. Even for the Italian Serie A in 2017-2018, the TRPS is similar for all methods after 30 match days,  although the RPS indicated that the predictions of the PFSC are worse for the remaining matches in that particular season. As for the RPS, the performances of the current-strength team ratings and the plus-minus player ratings are very similar.


\section{Discussion}\label{sec:France}

We have shown in the previous section that our simple PFSC is comparable in terms of predictive performance at match and final standing levels to the two benchmark models that are more computationally demanding, more time-consuming and require more input data. This establishes the PFSC as a very good candidate for obtaining fair final standings. 

We will now illustrate how our PFSC can be used in  practice by decision-makers to reach fairer decisions on how an abruptly stopped season should be evaluated. To this end, we will show how the French Ligue 1 could have been settled after it was abruptly stopped in the 2019-2020 season. The match results were downloaded from the website \url{https://www.football-data.co.uk/}. The code for the PFSC applied on the present case study is available in the online supplementary material. The official ranking of the Ligue 1 is given in Table \ref{table:official}.

\begin{table}[ht]
\caption{The official ranking of the French Ligue 1 in the 2019-2020 season. The order of the teams is determined by the number of points earned per match.}
\label{table:official}
\centering
\resizebox{\textwidth}{!}{\begin{tabular}{rlccccccccc}
  \hline
 & Team & Points & Win & Draw & Loss & Goals & \begin{tabular}{c}Goals\\ against\end{tabular} & \begin{tabular}{c}Goal\\ difference\end{tabular} & Matches & \begin{tabular}{c}Points\\per match\end{tabular} \\ 
  \hline
1 & PSG &  68 &  22 &   2 &   3 &  75 &  24 &  51 &  27 & 2.52 \\ 
  2 & Marseille &  56 &  16 &   8 &   4 &  41 &  29 &  12 &  28 & 2.00 \\ 
  3 & Rennes &  50 &  15 &   5 &   8 &  38 &  24 &  14 &  28 & 1.79 \\ 
  4 & Lille &  49 &  15 &   4 &   9 &  35 &  27 &   8 &  28 & 1.75 \\ 
  5 & Nice &  41 &  11 &   8 &   9 &  41 &  38 &   3 &  28 & 1.46 \\ 
  6 & Reims &  41 &  10 &  11 &   7 &  26 &  21 &   5 &  28 & 1.46 \\ 
  7 & Lyon &  40 &  11 &   7 &  10 &  42 &  27 &  15 &  28 & 1.43 \\ 
  8 & Montpellier &  40 &  11 &   7 &  10 &  35 &  34 &   1 &  28 & 1.43 \\ 
  9 & Monaco &  40 &  11 &   7 &  10 &  44 &  44 &   0 &  28 & 1.43 \\ 
  10 & Strasbourg &  38 &  11 &   5 &  11 &  32 &  32 &   0 &  27 & 1.41 \\ 
  11 & Angers &  39 &  11 &   6 &  11 &  28 &  33 &  -5 &  28 & 1.39 \\ 
  12 & Bordeaux &  37 &   9 &  10 &   9 &  40 &  34 &   6 &  28 & 1.32 \\ 
  13 & Nantes &  37 &  11 &   4 &  13 &  28 &  31 &  -3 &  28 & 1.32 \\ 
  14 & Brest &  34 &   8 &  10 &  10 &  34 &  37 &  -3 &  28 & 1.21 \\ 
  15 & Metz &  34 &   8 &  10 &  10 &  27 &  35 &  -8 &  28 & 1.21 \\ 
  16 & Dijon &  30 &   7 &   9 &  12 &  27 &  37 & -10 &  28 & 1.07 \\ 
  17 & \mbox{St.} Etienne &  30 &   8 &   6 &  14 &  29 &  45 & -16 &  28 & 1.07 \\ 
  18 & N\^imes &  27 &   7 &   6 &  15 &  29 &  44 & -15 &  28 & 0.96 \\ 
  19 & Amiens &  23 &   4 &  11 &  13 &  31 &  50 & -19 &  28 & 0.82 \\ 
  20 & Toulouse &  13 &   3 &   4 &  21 &  22 &  58 & -36 &  28 & 0.46 \\ 
   \hline
\end{tabular}}
\end{table}

\subsection{Evaluating the French Ligue 1 2019-2020 season}
At the time of stopping the competition, each team had played at least 27 matches. From the findings of the previous section, we know that after this number of match days, our PFSC is a competitive model for predicting the remaining matches and the final ranking. Based on the played matches, the teams strengths were estimated, which resulted in the strengths reported in Table \ref{table:ratings}. We can see that Paris Saint-Germain (PSG) is by far considered as the strongest team in the league. Surprisingly, Olympique Lyon comes out as the second strongest team, while only standing on the 7th place in the official ranking at that time. This could indicate that Lyon did have some bad luck during the season. Looking at their match results, we could see that in almost all their lost matches, they lost with a goal difference of 1 goal. Only PSG at home managed to get a margin of two goals against Lyon. At the bottom of the table, we find that Toulouse was the weakest team in the league, followed by Amiens, St. Etienne and N\^imes. This is in agreement with the official ranking,  up to a slightly different ordering of the teams.

\begin{table}
\caption{The estimated ratings $r_i, i=1,\ldots,18$, of the teams in the French Ligue 1, based on the played matches in the 2019-2020 season, obtained via the bivariate Poisson model in our PFSC.}
\label{table:ratings}
\centering
\begin{tabular}{rlr}
  \hline
 & Teams & Estimated strengths \\ 
  \hline
1 & PSG & 0.85 \\ 
  2 & Lyon & 0.28 \\ 
  3 & Rennes & 0.24 \\ 
  4 & Marseille & 0.19 \\ 
  5 & Lille & 0.15 \\ 
  6 & Bordeaux & 0.15 \\ 
  7 & Reims & 0.08 \\ 
  8 & Nice & 0.03 \\ 
  9 & Montpellier & 0.02 \\ 
  10 & Monaco & 0.01 \\ 
  11 & Strasbourg & $-$0.01 \\ 
  12 & Nantes & $-$0.02 \\ 
  13 & Brest & $-$0.09 \\ 
  14 & Angers & $-$0.10 \\ 
  15 & Metz & $-$0.16 \\ 
  16 & Dijon & $-$0.17 \\ 
  17 & N\^imes & $-$0.27 \\ 
  18 & \mbox{St.} Etienne & $-$0.29 \\ 
  19 & Amiens & $-$0.32 \\ 
  20 & Toulouse & $-$0.59 \\ 
   \hline
\end{tabular}
\end{table}

Using these strengths, we  have simulated the remainder of the season 100,000 times and by taking the mean over these simulations, we have calculated the probabilities for each team to reach each possible position, which is summarized in Table \ref{table:PFSC_France}. We can see that PSG would win the league with approximately $100\%$ probability, thanks to the big lead they had and the high team strength. Marseille had a $77\%$ chance of keeping the second position, with also a certain chance of becoming third ($18\%$), or even fourth ($5\%$). Furthermore, we see that Lyon, thanks to their high estimated strength, had the highest probability to be ranked as fifth ($29.5\%$). Their frustration with respect to the decision as it was taken officially by the Ligue 1 is thus understandable \citep{Holroyd2020}. In the bottom of the standing, we see that Toulouse was doomed to be relegated, with only $0.1\%$ chance of not ending at the 19th or 20th place in the league. Amiens had still about $32\%$ chance of staying in the first league.

\begin{table}[ht] \caption{The Probabilistic final standing (in percentages) of the Ligue 1 in the 2019-2020 season, according to our PFSC method. Probabilities are rounded to the nearest 1 percent.}
\label{table:PFSC_France}
\centering
\resizebox{\textwidth}{!}{\begin{tabular}{rllllllllllllllllllll}
  \hline
 & 1 & 2 & 3 & 4 & 5 & 6 & 7 & 8 & 9 & 10 & 11 & 12 & 13 & 14 & 15 & 16 & 17 & 18 & 19 & 20 \\ 
  \hline
PSG & 100 &  &  &  &  &  &  &  &  &  &  &  &  &  &  &  &  &  &  &  \\ 
  Marseille &  & 77 & 18 & 5 &  &  &  &  &  &  &  &  &  &  &  &  &  &  &  &  \\ 
  Rennes &  & 12 & 41 & 36 & 8 & 2 & 1 &  &  &  &  &  &  &  &  &  &  &  &  &  \\ 
  Lille &  & 11 & 37 & 39 & 9 & 3 & 1 &  &  &  &  &  &  &  &  &  &  &  &  &  \\ 
  Lyon &  &  & 3 & 10 & 30 & 18 & 13 & 9 & 6 & 4 & 3 & 2 & 1 &  &  &  &  &  &  &  \\ 
  Reims &  &  &  & 3 & 12 & 15 & 15 & 13 & 12 & 10 & 8 & 6 & 4 & 2 &  &  &  &  &  &  \\ 
  Montpellier &  &  &  & 3 & 11 & 13 & 13 & 12 & 12 & 10 & 9 & 7 & 5 & 3 & 1 &  &  &  &  &  \\ 
  Bordeaux &  &  &  & 2 & 7 & 11 & 12 & 12 & 12 & 11 & 10 & 9 & 7 & 4 & 1 &  &  &  &  &  \\ 
  Nice &  &  &  & 1 & 7 & 10 & 12 & 12 & 12 & 12 & 11 & 9 & 7 & 4 & 1 &  &  &  &  &  \\ 
  Strasbourg &  &  &  & 1 & 6 & 9 & 11 & 11 & 12 & 12 & 11 & 10 & 9 & 5 & 2 & 1 &  &  &  &  \\ 
  Monaco &  &  &  & 1 & 4 & 7 & 9 & 10 & 12 & 12 & 13 & 12 & 10 & 6 & 3 & 1 &  &  &  &  \\ 
  Nantes &  &  &  & 1 & 4 & 7 & 8 & 10 & 11 & 12 & 12 & 12 & 11 & 7 & 3 & 1 &  &  &  &  \\ 
  Angers &  &  &  &  & 2 & 4 & 5 & 7 & 9 & 11 & 13 & 15 & 15 & 11 & 5 & 2 & 1 &  &  &  \\ 
  Metz &  &  &  &  &  &  &  & 1 & 1 & 2 & 4 & 7 & 11 & 21 & 22 & 16 & 10 & 4 &  &  \\ 
  Brest &  &  &  &  &  &  &  & 1 & 1 & 2 & 4 & 6 & 10 & 19 & 23 & 18 & 10 & 5 & 1 &  \\ 
  Dijon &  &  &  &  &  &  &  &  &  &  & 1 & 2 & 4 & 10 & 16 & 22 & 23 & 15 & 5 &  \\ 
  \mbox{St.} Etienne &  &  &  &  &  &  &  &  &  &  & 1 & 1 & 3 & 7 & 14 & 22 & 25 & 20 & 7 &  \\ 
  N\^imes &  &  &  &  &  &  &  &  &  &  &  &  & 1 & 3 & 6 & 12 & 22 & 36 & 18 &  \\ 
  Amiens &  &  &  &  &  &  &  &  &  &  &  &  &  &  & 1 & 3 & 8 & 20 & 66 & 2 \\ 
  Toulouse &  &  &  &  &  &  &  &  &  &  &  &  &  &  &  &  &  &  & 2 & 98 \\ 
   \hline
\end{tabular}}

\end{table}

Now, how could this table be used by decision-makers to handle the discontinued season? One has to decide which team will become the champion, which teams will play in the Champions League and Europe League and which teams will be relegated to the second division.

For the first answer, some leagues nowadays have entered a rule stating that if enough matches are played, the current leader of the season would be considered as the champion. However, this does not take into account the gap between the first and the second in the standing. We would recommend changing the rule, in the sense that a team can only be declared champion if it has  more than $C\%$ chance to win the league according to the PFSC ($C$ could \mbox{e.g.} be $80$, but this decision  of course has to be made by the leagues). For our example, there is little doubt. PSG was expected to be the winner of the league with an estimated chance of $100\%$, so they should be considered as the champions of the Ligue 1. A similar strategy can be adopted regarding which teams should be relegated to the second division.

For the participation in the Champions League and Europe League, the leagues need a determined final standing instead of a probabilistic final standing. We will next show how we can get a determined final standing using our PFSC, and how we can use the PFSC to help to determine financial compensations.

\subsection{Determined final standing and financial compensations}
First we make a determined final standing by calculating the expected rank, using the probabilities. This results in the standing shown in Table \ref{table:determined}. In the example of the French League, we see that PSG gets the direct ticket for the Champions League, while Marseille and Rennes get the tickets for the CL qualification rounds. Lille and Lyon would have received the tickets for the group stage of the Europe League and Reims the ticket for the qualifications. This shows that Nice was one of the teams that got an advantage from the decision of the French league to halt the season.

However, transforming our probabilistic standing to a determined standing causes a number of teams to be (dis)advantaged. For example, in Table \ref{table:determined} we can see that the expected rank of Rennes is 3.55, which is the third-highest expected rank. Assigning Rennes to the third rank is therefore an advantageous outcome. Lille, on the other hand, has an expected rank of 3.62, which is only the fourth-best expected rank. Lille is therefore at a disadvantage when being assigned to rank 4. 

This issue could be solved by using a compensation fund. Assume that the expected profit (in particular prize money for the league placement and starting and prize money from Champions League and Europe League) of a team ending in rank $i$ is equal to $P_i$. The expected profit for \mbox{e.g.} Marseille would be $0.77*P_2+0.18*P_3+0.05*P_4$. In the determined ranking, they end as second, so they will receive $P_2$. Actually, they receive too much, since they had no chance of ranking higher than second, but they had a reasonable chance to become third or even fourth. To compensate for this, they should hand over $P_2-(0.77*P_2+0.18*P_3+0.05*P_4)=0.18*(P_2-P_3)+0.05*(P_2-P_4)$ to the compensation fund. This will then be used to compensate teams that are disadvantaged by the establishing of a determined ranking. There will still be the difficulty of estimating the expected profit from reaching a certain rank (\mbox{e.g.}, a team reaching the Europe League will have further merchandising advantages besides the profit mentioned above as compared to the team classified just outside of these ranks), but we believe that this tool could be very useful for decision-makers in determining which teams have received an advantage or disadvantage from an early stop of the league, and how to compensate for this.

\begin{table}[ht]
\caption{Determined final standing, using the PFSC probabilities. This standing could be used to decide which teams will play in the Champions League and Europe League.}
\label{table:determined}
\centering
\begin{tabular}{rlr}
  \hline
Rank & Team & Expected rank \\ 
  \hline
1 & PSG & 1.00 \\ 
  2 & Marseille & 2.28 \\ 
  3 & Rennes & 3.55 \\ 
  4 & Lille & 3.62 \\ 
  5 & Lyon & 6.52 \\ 
  6 & Reims & 8.17 \\ 
  7 & Montpellier & 8.51 \\ 
  8 & Bordeaux & 9.11 \\ 
  9 & Nice & 9.20 \\ 
  10 & Strasbourg & 9.58 \\ 
  11 & Monaco & 10.06 \\ 
  12 & Nantes & 10.17 \\ 
  13 & Angers & 11.08 \\ 
  14 & Metz & 14.38 \\ 
  15 & Brest & 14.53 \\ 
  16 & Dijon & 16.00 \\ 
  17 & \mbox{St.} Etienne & 16.40 \\ 
  18 & N\^imes & 17.34 \\ 
  19 & Amiens & 18.52 \\ 
  20 & Toulouse & 19.98 \\ 
   \hline
\end{tabular}
\end{table}

\section{Conclusion}\label{sec:final}

In this paper we proposed a novel tool, the Probabilistic Final Standing Calculator, to determine the most likely outcome of an abruptly stopped football season. Unlike other recent proposals, we provide probabilities for every team to reach each possible rank, which is more informative than only providing the single most likely or expected final ranking. We have shown the strength of our PFSC by comparing it to two benchmark models that are based on much more information, and yet our PFSC is exhibiting similar performances except when a season would get stopped extremely early, which  however was anyway more a theoretical than a practical concern (a season stopped after less than a third of the games played would certainly be declared void). Our evaluation has been done at both the match-level (via the RPS) and the final standing level (via the TRPS).

We have shown on the concrete example of the 2019-2020 season of the French Ligue 1 how our PFSC can be used, also for a fair division of the money shares. We hope that our PFSC will help decision-makers in the future to reach fairer decisions that will not lead to the same level of dissatisfaction and controversies that one could observe in various countries in the 2019-2020 season. The idea of the PFSC can also be applied, up to minor modifications, to several other types of sports tournaments.

We conclude this paper with a historical remark. The problem treated here, namely finding a fair way to determine final league standings if one cannot continue playing, goes back to the very roots of probability theory. The French writer Antoine Gombaud (1607-1684), famously known as Chevalier de M\'er\'e, was a gambler interested by the ``problem of the points'': if two players play a pre-determined series of games, say 13, and they get interrupted at a score of, say, 5-2, and cannot resume the games, how should the stake be divided among them? The Chevalier de M\'er\'e  posed this problem around 1654 to the mathematical community, and the two famous mathematicians Blaise Pascal (1623-1662) and Pierre de Fermat (1607-1665) accepted the challenge. They addressed the problem in an exchange of letters that established the basis of our modern probability theory \citep{Dev10}.

\section{Declarations}
Funding: The authors have no funding to report\\
Conflicts of interest/Competing interests: The authors have no conflict of interest to declare.\\
Availability of data and material: Data needed for the application\\
Code availability: The code for the application is available in a supplementary file.

\bibliography{pm}

\end{document}